\documentclass[twocolumn,showpacs,preprintnumbers,amsmath,amssymb,prl] {revtex4}
\usepackage{graphicx}
\usepackage{dcolumn}
\usepackage{bm}
\usepackage{eucal}

\begin{document}

\title{Non-Unitary Spin-Charge Separation in One-Dimensional Fermion Gas}

\author{Vadim~V.~Cheianov$^{1}$ and M.B.~Zvonarev$^{2,3}$}
\affiliation{$^1$NORDITA, Blegdamsvej 17, Copenhagen {\O},  DK 2100, Denmark
\\ $^2$\O rsted Laboratory, Niels Bohr Institute for APG, Universitetsparken 5, Copenhagen {\O}, DK 2100, Denmark
\\
$^3$Petersburg Department of Steklov Institute of Mathematics,
Fontanka 27, St~Petersburg 191011, Russia
}

\date{\today}

\begin{abstract}
In this Letter we report exact results on the infrared asymptotics
of one-particle dynamical correlation function of the  gas of
impenetrable spin 1/2 fermions at infinitesimal temperature.  The
correlation function shows signs of spin-charge separation with
scaling behavior in the charge part and exponential decay as a
function of the space coordinate in the spin part. Surprisingly,
the anomalous dimensions in the charge part do not correspond to
any unitary conformal field theory. We find that the fermion
spectral weight has a power law divergency at low energy with the
anomalous exponent $ -1/2.$
\end{abstract}


\maketitle Significant advance of the theory of strongly
correlated one dimensional electron systems has been made possible
by two important peculiarities of one dimensional physics:
universality of low energy properties summarized by the conformal
field theory (CFT) \cite{Tsvelik} and the existence of exactly
solvable strongly coupled models like the Hubbard model or the
Thirring model \cite{KBI}. As far as their low energy properties
are concerned, most physically important one dimensional electron
systems belong to a certain universality class usually referred to
as the Luttinger liquid (LL) \cite{Haldane}. By bosonization the
LL theory is mapped onto a theory of free massless compactified
bosonic fields. The compactification radii play the role of
phenomenological parameters defining the effective theory.
Classification and phenomenology of such effective theories is
often called g-ology \cite{Solyom}. Due to a relatively simple
analytic structure of the LL theory,  g-ology is a powerful
calculational tool relating all dynamical information about a
strongly correlated system (such as its correlation functions,
spectral weights,  transport coefficients, stability with respect
to perturbations etc.) to a few phenomenological constants, which
can be obtained from e.g. its thermodynamical properties. For
exactly solvable models this phenomenological constants are
usually found from the analysis of thermodynamics by means of the
thermodynamic Bethe ansatz \cite{Haldane1}.

Despite the success of g-ology, a direct  calculation of infrared
asymptotics of correlation functions in exactly solvable models
has so far been performed only for a few special cases
\cite{Berkovich,Ogata,Parola,Ha,Gohmann,KBI} and as a general
problem still remains a challenge. Such direct calculations are
important for two reasons. First, they provide a test bed for the
hypotheses underlying the LL theory. Second, they give insight
into the situations where the low energy physics is not described
by the LL.

An interesting example of a physically relevant model, where exact
calculation of dynamical correlation functions is possible is a
model of non-relativistic  impenetrable spin 1/2  fermions. This
model is the infinite $U$ limit of the system of spin 1/2 fermions
described by the Hamiltonian
\begin{equation}
H=\int d x \left[ -\sum_{\sigma =\uparrow, \downarrow}\psi^\dagger_\sigma(x)
\partial_x^2 \psi_\sigma(x)+ U n_\downarrow (x) n_\uparrow(x)\right ].
\label{Hamiltonian}
\end{equation}
The model \eqref{Hamiltonian} can be viewed as a continuum limit
of the Hubbard model and is the simplest example of interacting
fermions in one dimension. This model has a long history. The
eigenstates and the spectrum of Hamiltonian \eqref{Hamiltonian}
for an arbitrary $U$ were found in \cite{Yang}. Thermodynamics of
this model was studied in \cite{Lai}. Parameters of the low energy
LL theory for the Hubbard model were calculated in \cite{Korepin}.

The  infinite $U$  limit of the model was studied in
\cite{Ogata,Shiba,Parola,Gohmann,Anderson,Berkovich,Izergin}. This
is an interesting limit, where the ground state of the system
becomes infinitely degenerate in spin  and the CFT description
fails.  On the other hand, in this limit a specific factorization
of Yang's wave function occurs \cite{Ogata}, opening the way to a
direct calculation of spectral weights and correlation functions
of the model through their Fredholm determinant representation
\cite{Berkovich,Izergin}. Due to the above mentioned ground state
degeneracy the limit $U\to \infty$ in correlation functions of the
model is nontrivial in the sense that it does not commute with the
limit of vanishing temperature $T\to 0$. Indeed, taking the limit
$T\to 0$ first one obtains results \cite{Ogata,Parola,Shiba}
consistent with a naive infinite $U$ limit of the corresponding
CFT  \cite{KBI,Anderson}. In particular,  the one-particle
momentum distribution function shows scaling behavior near the
Fermi momentum $k_F$ with the anomalous exponent $1/8$
\cite{Ogata}. On the other hand, the one-particle density matrix
(i.e. the equal time one-particle correlation function) decays
exponentially as a function of coordinate \cite{Berkovich} if one
takes the $U\to\infty$ limit first. Due to this exponential decay
the momentum distribution function shows no scaling near $k_F.$

In this Letter we report our results on the asymptotics of the
one-particle dynamical correlation function
\begin{equation}
G(x,t)=\left\langle \psi^{\dagger}_\uparrow(x,t)\psi_\uparrow (0,0) \right\rangle
\label{corf}
\end{equation}
when the  limit  $U\to \infty$ is taken before the limit $T\to 0$.
Due to the rotational invariance of Hamiltonian
\eqref{Hamiltonian} the correlation function of spin down fermions
coincides with \eqref{corf}.

We begin our discussion with first recalling how the infinite $U$
limit in correlation functions is taken at zero temperature and
then we discuss the opposite order of limits qualitatively. Then
we present our exact result for the asymptotics of the correlation
function \eqref{corf} and discuss its structure and physical
meaning.

Assume that in the ground state of the system there  is  a  finite
fermion density $n_c=n_\uparrow+n_\downarrow,$ controlled by the
chemical potential $\mu.$ At any finite $U$ the low energy physics
of this model can be described by the LL theory. According to this
theory the low energy spectrum is described by the effective
Hamiltonian
\begin{equation}
H_{\rm eff}=H_c+H_s
\label{Heff}
\end{equation}
where $H_c$ describes charge fluctuations, $H_s$ describes spin
fluctuations and $[H_c, H_s]=0.$  Independent dynamics of spin and
charge is called spin-charge separation \cite{Tsvelik}.
Hamiltonians $H_c$ and $H_s$ are, in fact, Tomonaga-Luttinger
Hamiltonians corresponding to $c=1$ conformal field theories with
global $U(1)$ symmetry. Hamiltonian $H_s$ also possess global
$SU(2)$ symmetry inherited from the rotational symmetry of
\eqref{Hamiltonian}. This additional constraint on $H_s$ fixes the
structure of its spectrum completely up to a non-universal
constant $v_s$, which is the spin propagation velocity depending
on the microscopic parameters. The fermion operator in the LL
theory is represented as a sum of  anticommuting ``left'' and
``right''  fermion fields
\begin{equation}
\psi_\sigma(x) = \psi_{L, \sigma}(x) e^{-i k_F x} + \psi_{R, \sigma}(x) e^{i k_F x},\quad \sigma=\uparrow,\downarrow
\label{fermion}
\end{equation}
where $k_F$ is the Fermi momentum. The ``left'' and ``right''
fermion fields are products of spin and charge vertex operators
\begin{equation}
\psi_{L, \sigma}=  C_L  S_{L, \sigma} , \quad \psi_{R, \sigma}=  C_R  S_{R, \sigma},
\label{psisplit}
\end{equation}
where the spin operators satisfy $ S_{L(R),\downarrow}=S_{L(R),\uparrow}^\dagger.$
Operators $S$ and $C$ commute with each other and satisfy $[S, H_c]=[C, H_s]=0.$
Formulas \eqref{Heff}, \eqref{fermion} and \eqref{psisplit} only apply to the low
energy properties of the system. That is, the energy must be much smaller than a certain
cutoff scale $\Lambda,$ determined by $\mu$ and $U.$

At zero temperature, the correlation function \eqref{corf}
predicted by the Luttinger theory has the form
\begin{align}
G(x,t)=\mathcal S(x-v_s t) \mathcal C(x-v_c t, x+v_c t) e^{-i k_F x}
\nonumber
\\
+\mathcal S(x+v_s t) \mathcal C(x+v_c t, x-v_c t) e^{i k_F x}
\label{LLcorf}
\end{align}
where
\begin{equation}
\mathcal S(x-v_s t)= \left \langle S_R^\dagger (x,t) S_R(0,0) \right \rangle = \frac{\mathrm{const} }{(x-v_s t)^{1/2}}
\label{Scorrelator}
\end{equation}
and
\begin{align}
\mathcal C(x-v_c t, x+v_c t)&=
\left\langle C_R^{\dagger}(x,t) C_R(0,0) \right\rangle
\nonumber
\\
&=
\frac{\mathrm{ const}}{(x-v_c t)^{2 \Delta_c} (x+v_c t)^{2\bar \Delta_c}},
\label{Ccorrelator}
\end{align}
where $v_s$ and $v_c$ are the propagation velocities of spin and
charge excitations respectively. Correlation functions
\eqref{Scorrelator}, \eqref{Ccorrelator} do not contain any length
parameter. This reflects the scaling invariance of the LL theory.
According to \eqref{Scorrelator} the spin operator $S_{R,\sigma}$
is a right mover with fixed anomalous dimension 1/4. This is a
consequence of the constraint imposed on the effective Hamiltonian
$H_s$ by the global $SU(2)$ symmetry. At the same time, the
correlation function of the "right" charge operator $C_R$  splits
into a product of left-moving (depending on $x+v_c t$) and
right-moving (depending on $x-v_c t$) parts with different
positive anomalous dimensions $\Delta$ and $\bar \Delta.$ These
dimensions are determined by  the Luttinger parameter $K,$ which
can be found from the charge compressibility $\kappa=\partial
n_c/\partial \mu$ of the system as follows
\begin{equation}
K=\frac{\pi v_c}{2}\kappa.
\label{Kdef}
\end{equation}
One has \cite{Tsvelik}
\begin{align}
\Delta_c&=\frac{1}{16}\left(\sqrt K+ \frac{1}{\sqrt K} \right)^2 \quad,
\nonumber
\\
 \bar \Delta_c&=\frac{1}{16}\left(  \sqrt K- \frac{1}{\sqrt K}    \right)^2.
\label{LLdimensions}
\end{align}
For any finite $U$ parameters $v_c$ and $\kappa$ can be calculated
from the thermodynamic Bethe ansatz.  For infinite $U$ fermions
become impenetrable, which from the point of view of
thermodynamics means that they satisfy the Pauli principle
independently of their spin orientation. For such a system
parameters $v_c$ and $\kappa$ are calculated in the same way as
for a system of spinless non-interacting fermions
\cite{Shultz,Anderson}. Calculating the right hand side of
\eqref{Kdef} for spinless fermions one finds
\begin{equation}
K=\frac{1}{2}.
\label{Kinf}
\end{equation}
For the anomalous dimensions in the charge sector this gives
\begin{equation}
\Delta_{c}=\frac{9}{32}, \quad \bar \Delta_{c}=\frac{1}{32}.
\label{naivedimensions}
\end{equation}
Correlation function for impenetrable fermions  in the form \eqref{LLcorf} with
anomalous exponents in the charge sector given by \eqref{naivedimensions}
was suggested in \cite{KBI}. It was also discussed in \cite{Shiba}, where the
$U\to\infty$ limit was analyzed microscopically.

Next, consider the infinite $U$ limit taken at finite temperature.
For large $U$ there happens a separation of energy scales in the
problem. This separation is controlled by a small dimensionless
parameter $k_F/U.$ While $\mu$ remains the characteristic cutoff
energy  in the charge sector, the characteristic cutoff energy in
the spin sector becomes of the order of $ \mu k_F/U.$ The same
happens with the velocities. While the propagation velocity for
the charge $v_c\to 2 \sqrt \mu$ as $U\to \infty,$ the velocity of
spin excitations vanishes, $v_s\to 0.$ In the limit of infinite
$U$ the spin degrees of freedom completely loose dynamics and the
ground state of the system becomes infinitely degenerate with
respect to spin flips. From the point of view of the low energy
physics it is interesting to consider a situation where
\begin{equation}
\frac{\mu k_F}{U}\ll T \ll \mu
\label{temperature}
\end{equation}
In this situation all spin configurations have equal statistical
weight and the dynamics of the system is effectively averaged over
spin configurations. At the same time, the charge degrees of
freedom are not strongly affected by temperature, because the
latter is much lower than the corresponding cutoff energy $\mu.$
Thus in the temperature range \eqref{temperature} correlation
function \eqref{corf} should not strongly depend on temperature
and its behavior should be in this sense universal. The infinite
$U$ limit of the correlation function taken at a fixed small
temperature $T\ll\mu$  thus corresponds to taking the infinite $U$
limit first and then taking the limit $T\to0.$ One can try to
conjecture the structure of the one-particle correlation function
\eqref{corf} in this situation using formulas \eqref{Scorrelator},
\eqref{Ccorrelator} and the general results of the CFT. For finite
temperatures and for $x> v_s/T$, the spin part  of the correlation
function predicted by the LL theory is
\begin{equation}
 \mathcal S_R(x-v_s t)\sim  e^{- \frac{2\pi T}{v_s}(x- v_s t)}.
\label{TScorrelator}
\end{equation}
This formula is valid for temperatures smaller than $\Lambda_s=\mu
k_F/U\approx k_F v_s,$ that is when linear description of spin
waves is appropriate. For $\vert x\pm v_c t\vert <v_c/T$  the
charge part of correlation function is given by
\eqref{Ccorrelator}. For $T>\Lambda_s$ all spin degrees of freedom
become saturated and the spin part of the correlation function
should become temperature independent. Thus, replacing the
temperature $T$ in the right hand side of \eqref{TScorrelator} by
the crossover scale $\Lambda_s$ we get a prediction for the
correlation function at $T>\Lambda_s.$ Making this substitution
and taking  the limit of vanishing $v_s$ we get for the fermion
correlation function
\begin{eqnarray}
G(x,t)=\mathcal S(x)\mathcal C(x+v_c t, x-v_c t) e^{-i k_F x}
\nonumber
\\+
\mathcal S(x)\mathcal C(x-v_c t, x+v_c t) e^{i k_F x},
\label{conjecture}
\end{eqnarray}
where function $\mathcal C$ is given by \eqref{Ccorrelator} and
\begin{equation}
\mathcal S(x)={\rm const}\times e^{-\gamma k_F x}.
\label{LimitS}
\end{equation}
Here $\gamma$ is a dimensionless constant.
The problem of this picture is that it assumes spin-charge
separation in the LL sense even though the spin degrees are
strongly excited. To see that this is not the case, compare the
correlation function of spin up fermions calculated in two
different  ground states. Let one ground state be the infinite $U$
limit of the ground state of Hamiltonian \eqref{Hamiltonian} and
another be the fully polarized spin up ground state. Both states
have the same statistical weight and equally contribute to the
correlation function. In the first state the anomalous dimensions
of charge part of fermion correlation function are given by
\eqref{naivedimensions}. In the second state, due to the Pauli
principle fermions do not feel the interaction and the anomalous
dimensions are those of non-interacting fermions, that is
$\Delta=1/2$ and $\bar \Delta=0.$ It is not, therefore, clear
whether after averaging over all spin configurations the anomalous
dimensions in \eqref{conjecture} should coincide with
\eqref{naivedimensions} or, even, whether the correlation function
should have the structure \eqref{conjecture} at all.

In paper \cite{Izergin} it was shown that the correlation function
\eqref{corf} can be expressed in terms of the Fredholm determinant
of a linear integral operator $\hat V$ via
\begin{equation}
G(x,t)=\frac{e^{-i k_F^2 t}}{8\pi i} \oint_{|z|=1} \frac { d z}
{z} F(z) B_{--}(z)\det(\hat I+\hat V)(z). \label{Gintz}
\end{equation}
The kernel
\begin{equation}
V(k,p)=\frac{e_+(k)e_-(p)-e_+(p)e_-(k)}{k-p}
\label{Vdef}
\end{equation}
of the operator $\hat V$ is defined on the square
$[-k_F,k_F]\times[-k_F, k_F],$ where $k_F=\sqrt \mu.$ Functions
entering \eqref{Vdef} are defined as follows ($z=e^{i\eta}$)
\begin{align}
& e_-(k)=\frac{1}{\sqrt \pi} e^{\tau(k)/2}, \\
& e_+(k)=\frac{e^{-\tau(k)/2}}{2\sqrt\pi} [(1-\cos\eta)e^{\tau(k)} E_0(k)+\sin\eta] ,\\
&E_0(k)=\text{p.v.} \int_{-\infty}^{\infty} d p \frac{e^{-\tau(p)}}{\pi(p - k )},\\
&\tau(k) = i k^2 t- i k x.
\end{align}
Function $B_{--}(z)$ is
\begin{equation}
B_{--}(z) = \int_{-k_F}^{k_F} d k e_{-}(k) (\hat I + \hat V)^{-1} e_{-}(k)
\label{Bmm}
\end{equation}
and
\begin{equation}
\label{Fdef}
F(z)=1+\frac{z}{2-z}+\frac{1}{2 z-1}.
\end{equation}

We performed asymptotic analysis of Eq. \eqref{Gintz} by asymptotically solving the
corresponding Riemann-Hilbert problem \cite{Gohmann} by the techniques described in
\cite{Deift}. Detailed calculations are involved and will be presented
elsewhere \cite{Us}.
Our main results for the correlation function \eqref{corf} are as follows.
For $x, t \to +\infty$ and $x/t=\mathrm{const} \neq 0$ we have
\begin{align}
G(x,t)=&\frac{\Xi e^{-k_F x \ln 2 / \pi }
e^{i(k_F x-\phi_+) }  }{(x-2 k_Ft)^{2 \Delta}(x+2k_Ft)^{2 \bar \Delta} }
\nonumber
\\
&
-\frac{\Xi e^{-k_F x \ln 2 / \pi } e^{-i(k_F x-\phi_- )}
}{(x+2 k_F t)^{2 \Delta}(x-2 k_F t)^{2 \bar \Delta} }.
\label{answer}
\end{align}
Here  $\Xi$ is a constant, which is calculated explicitly in \cite{Us}.
The phases $\phi_{\pm}$ are given by
\begin{equation}
\phi_{\pm}=-2{ \rm Im}\left[ \ln\Gamma\left(
\frac{i \ln 2}{2\pi}\right)\right]+\frac{\ln 2}{\pi}\ln(2 k_F x\pm 4k_F^2 t)
\label{phidef}
\end{equation}
and the anomalous dimensions are given by
\begin{equation}
\Delta= \frac{1}{2}- \frac{1}{8} \left(\frac{\ln 2}{\pi}\right)^2, \qquad
\bar \Delta=- \frac{1}{8} \left(\frac{\ln 2}{\pi}\right)^2.
\label{dimensions}
\end{equation}
This result \eqref{answer} is consistent with the previously
calculated equal time correlation function \cite{Berkovich}. The
case $x=0$ is special. In this case
\begin{equation}
G(0,t)= \frac{\Xi'}{\sqrt {t\ln (k_F^2 t)}} \label{zeroX}
\end{equation}
where $\Xi'$ is a constant.

The form of the correlation function \eqref{answer} is essentially
the same as of \eqref{conjecture} and shows clear signs of
spin-charge separation. The exponentially decaying part can be
attributed to the non-propagating spin mode. The algebraically
decaying part of the correlation function only depends on
combinations $x-2 k_F t$ and $x+2 k_F t$ which correspond to
propagation with velocity $2 k_F.$ This is exactly the velocity of
the charge propagation in the model.

Remarkably, the anomalous dimensions given by \eqref{dimensions}
are completely different from dimensions \eqref{naivedimensions},
obtained in the infinite $U$ limit of the LL theory. While the
form of the correlation function \eqref{answer} suggests that the
fermion operator can be represented as a product of a spin
operator and a charge operator, the emergence of a negative
scaling dimension $\bar \Delta$ in \eqref{dimensions} indicates
that the conformal field theory describing the charge sector
should be  non-unitary. Answering the question of whether such a
CFT exists is a subject of further investigation. In particular,
it would be interesting to calculate the three point correlation
functions involving two fermion operators and the charge density
or the spin density operator.

Note that the phases $\phi_{\pm}$ \eqref{phidef} instead of being
constant are logarithmic functions of the light cone coordinates
$x\pm v_c t.$ A similar phenomenon occurs in the theory of
classical non-linear wave equations\cite{Manakov}.

While the exponential decay of the correlation function in $x$
direction smears the power law scaling of the momentum
distribution function near the Fermi points, the power law scaling
can be observed in the tunnelling density of states, which is
given by
\begin{equation}
A(\omega)=\text{Re} \tilde  G(0, \omega)
\end{equation}
where tilde denotes the Fourier transform. From our result \eqref{zeroX}
we find
\begin{equation}
A(\omega)\sim \omega^{-\frac{1}{2}}.
\end{equation}
The divergence of the tunnelling density of states at zero energy
is similar to the one found in \cite{Shiba} but the numerical
value of the tunnelling exponent $1/2$ is different from $3/8$
calculated in \cite{Shiba}.

M.B. Zvonarev's work was supported by the Danish Technical Research Council via the
Framework Programme on Superconductivity.

\end{document}